\documentclass{article}
\pdfoutput=1
\usepackage{amsmath,amssymb,amsfonts}
\usepackage{graphicx,color}
\usepackage{feynmp}
\usepackage{calc}
\usepackage{multicol}
\usepackage[hyperindex,breaklinks]{hyperref}
\usepackage[numeric,initials]{amsrefs}
\usepackage[width=372pt,height=588pt,top=126pt]{geometry} 

	\newcommand{\del}{\partial}
	\newcommand{\eps}{\epsilon}
\renewcommand{\d}{\mathrm{d}}
  \newcommand{\oo}{\infty}

\title{Comment on `Hawking radiation from fluctuating black holes'}
\author{Igor Khavkine \\
	\small
	Institute for Theoretical Physics,\\
	\small
	Utrecht University, Leuvenlaan 4, NL-3584 CE Utrecht, The Netherlands\\
	\small
	E-mail: i.khavkine@uu.nl}

\begin{document}\setlength{\unitlength}{1mm}
\maketitle

\begin{abstract}
	\noindent
	Takahashi \& Soda (2010 \textit{Class.\ Quantum Grav.}\ \textbf{27}
	175008) have recently considered the effect (at lowest non-trivial
	order) of dynamical, quantized gravitational fluctuations on the
	spectrum of scalar Hawking radiation from a collapsing Schwarzschild
	black hole. However, due to an unfortunate choice of gauge, the
	dominant (even divergent) contribution to the coefficient of the
	spectrum correction that they identify is a pure gauge artifact.
	I summarize the logic of their calculation, comment on
	the divergences encountered in its course and comment on how
	they could be eliminated, and thus the calculation be completed.
	\vspace{1ex}


\end{abstract}

\section{Introduction}
In their recent work~\cite{TS}, Takahashi~\& Soda have tackled an
interesting and challenging question, that of assessing the influence of
dynamical, quantized metric fluctuations on the spectrum of scalar Hawking
radiation from a black hole. A natural sibling question that could be
directly attacked with essentially the same techniques is that
of back reaction of Hawking radiation on the quantum geometry of the
black hole. An answer to either of these questions would provide us
with valuable insight into the properties of quantum gravity, as seen
through the prism of its reduction to effective quantum field theory on
a curved background.

The authors of~\cite{TS} show that the spectrum of scalar Hawking
radiation from a collapsing Schwarzschild black hole is corrected due to the
presence of dynamical, quantized gravitational fluctuations interacting
with the quantum scalar field at the lowest non-trivial (cubic) order.
The choice of a collapsing spacetime instead of an eternal black hole
is similar to the choice made in Hawking's original
calculation~\cite{H-bhrad}. The same result could be obtained by using
the Unruh vacuum on an eternal black hole spacetime~\cite{U-vac}.
The corrected expected number of quanta (expected by an asymptotic
observer at future null infinity) in a single field mode of static
frequency $\omega$ and spherical harmonic index $\ell$, according to
equations~(71), (81) and~(169) of~\cite{TS}, is
\begin{equation}\label{spec-corr}
	\langle N_{\omega\ell} \rangle =
		\frac{1}{e^{2\pi\omega/\kappa}-1}
		+ C_\ell \frac{\eps^2\omega_{\mathrm{cut}}^3\ell_p^2}{\kappa^3 M^2 L}
			\coth(\pi\omega/\kappa) + \cdots,
\end{equation}
where $C_\ell$ is a numerical constant, $M$ is the Schwarzschild mass,
$\kappa=1/(4M)$ is the surface gravity, $\ell_p$ is the Planck length,
$\eps>0$ is a small distance cutting off radial integration before the
horizon, and $\omega_{\mathrm{cut}}$ and $L$ are parameters regulating
the divergence of a 1-loop Feynman integral. Higher order corrections
are presumed to be subleading in $\ell_p$ or one of the regulator
parameters.

I will summarize some of they key steps in the calculation leading
to~\eqref{spec-corr} and discuss the divergences that appear along the
way, as well as ways of resolving them.  Section~\ref{sec:bg} reviews
some relevant background material and establishes the notation.
Section~\ref{sec:gauge} discusses the divergence that prompted the
introduction of the $\eps$-regulator, how it is related to the authors'
choice of gauge and how a better choice of gauge eliminates this
divergence and the need for this regulator. Section~\ref{sec:loop}
discusses the divergence that prompted the introduction of the
$\omega_{\mathrm{cut}}$ and $L$ regulators and how these regulators
could be removed using standard perturbative renormalization. Finally,
section~\ref{sec:discuss} emphasizes that, as a consequence of the
results of section~\ref{sec:gauge}, the dominant contribution,
identified in~\cite{TS}, to the coefficient in front of the correction
in equation~\eqref{spec-corr}, is a pure gauge artifact. It also
summarizes the non-trivial steps achieved in the calculation of
Takahashi~\& Soda and how it could be completed to obtain a
reliable, parameter-free estimate of the size of the correction to the
Hawking spectrum.

\section{Background and notation}\label{sec:bg}
The calculation in~\cite{TS} starts out by quantizing free scalar and
metric perturbation (graviton) fields on a collapsing Schwarzschild
background. The Bogolubov coefficients, which transform between
asymptotic in- and out-modes, are estimated using a geometric optics
approximation and give the standard Hawking spectrum for the free scalar
field. Then, the metric perturbations are gauge fixed and reduced to two
physical scalar degrees of freedom, which are quantized as free
fields. Finally, the explicit form of the interaction vertex between the
scalar field and the metric perturbations is identified and used to compute
the correction to the Hawking spectrum of the scalar field at lowest
perturbative order. The relevant details of these steps are discussed in
the rest of this comment.

An important
step in quantizing a free field is solving its classical equations of
motion. Usually, this is accomplished by decomposing an arbitrary
solution into a set of modes. On a Minkowski background, due to the
translational symmetry, it is easiest to work with the set where each
mode function is proportional to a plane wave
\begin{equation}
	\phi_k^0(x) = \exp(ik_\mu x^\mu),
\end{equation}
where $k_\mu$ is the wave vector indexing the mode and $x^\mu$ are
global inertial coordinates. We restrict ourselves only to the case of
massless fields, hence $k^2=0$. On a curved background, we can no longer
make use of translation symmetry. However, the background used
in~\cite{TS} is presumed to be (at least outside the horizon)
spherically symmetric, static in the asymptotic future, and
approximately static in the asymptotic past. To make maximal use of the
available symmetry, it is convenient to decompose the fields into a set
of modes, where each mode function is proportional to a scalar or tensor
spherical harmonic as well as $\phi^{+}_{\omega\ell j}(t,r)$ or
$\phi^{-}_{\omega\ell j}(t,r)$, where $j$ indexes independent tensor
polarizations (if any). These functions have the following asymptotic
properties
\begin{equation}
	\phi^\pm_{\omega\ell j} \sim e^{-i\omega t} R^\pm_{\omega\ell j}(r)
	\quad \text{for $t\to\pm\oo$},
\end{equation}
where, up to normalization, the radial functions $R^\pm_{\omega\ell j}$
are uniquely specified by the geometry and the above listed conditions.
The fact that the $\phi^-_{\omega\ell j}$ and $\phi^+_{\omega\ell j}$
mode functions do not coincide is ultimately responsible for the Hawking
effect~\cite{FN-bh}*{Sec.10.2}.

While the decomposition of a scalar field into spherical harmonics is a
standard exercise, the case of metric perturbations is more subtle, but
is a topic with an extensive literature, starting with the seminal work
of Regge \& Wheeler~\cite{RW-pert}. An up to date review can be found
in~\cite{BCS-quasi}. Reference~\cite{MP-pert} is particularly useful as
it presents the formalism of metric perturbations in a spherically
symmetric spacetime in a way that is gauge invariant and covariant with
respect to changes of coordinates in the $(t,r)$-plane. For reference,
we establish the correspondence between the notations of~\cite{TS}
and~\cite{MP-pert}.

Due to spatial inversion symmetry, the perturbations naturally decompose
into odd and even parity sectors. We shall only consider the even parity
ones. Note that, below, asterisks denote components that can be deduced
from symmetry.
\begin{align}
	\text{Takahashi \& Soda~\cite{TS}:} ~~
	h_{\mu\nu} &= \begin{pmatrix}
			f\bar{H} & H_1 & v_{|a} \\
			* & H/f & w_{|a} \\
			* & * & r^2 K\gamma_{ab} + B_{|ab}
		\end{pmatrix}
\end{align}
The components of the metric perturbation $h_{\mu\nu}$ are given in
static Schwarzschild coordinates%
	\footnote{More precisely, as described previously, they are
	approximately static coordinates on the collapsing spacetime.}, %
$f(r)=1-2M/r$, $\gamma_{ab}$ is the standard metric on the unit 2-sphere,
and $T^{\cdots}_{\cdots|a\cdots}$ denotes covariant differentiation
of the tensor $T$ with respect to $\gamma_{ab}$. The components of the
perturbation are parametrized by the scalars $\bar{H}$, $H_1$, $H$, $v$,
$w$, $K$ and $B$. The components $f\bar{H}$ and $H/f$ correspond to
$h_{tt}$ and $h_{rr}$ respectively.
\begin{align}
	\text{Martel \& Poisson~\cite{MP-pert}:} ~~
	p_{\mu\nu} &= \sum_{\ell m} \begin{pmatrix}
			h^{m}_{ab} Y^{\ell m} & j^{\ell m}_a Y^{\ell m}_B \\
			* & r^2 K^{\ell m} \Omega_{AB} Y^{\ell m} + r^2 G^{\ell m} Y^{\ell m}_{AB}
		\end{pmatrix}
\end{align}
The components of the metric perturbation $p_{\mu\nu}$ are given with
respect to coordinates that respect spherical symmetry but are arbitrary
on the $(t,r)$-plane, $\ell$ and $m$ are respectively the orbital and
magnetic spherical harmonic indices and $\Omega_{AB}$ is the standard
metric on the unit 2-sphere. The components of the perturbation are
parametrized by the scalars $K^{\ell m}$, $G^{\ell m}$, and the
$(t,r)$-plane tensors $h^{\ell m}_{ab}$ and $j^{\ell m}_a$. The vector
and tensor spherical harmonics are defined from the scalar $Y^{\ell m}$
as follows: $Y^{\ell m}_A = D_A Y^{\ell m}$ and $Y^{\ell m}_{AB}=[D_A
D_B+\frac{1}{2}\ell(\ell+1)\Omega_{AB}] Y^{\ell m}$, where $D_A$ is the
covariant derivative with respect to $\Omega_{AB}$. Where no confusion
is possible, the spherical harmonics indices may be omitted.

The correspondence between the two notations should now be clear, given
the equality $h_{\mu\nu}\d{x}^\mu\d{x}^\nu =
p_{\mu\nu}\d{x}^\mu\d{x}^\nu$. Though, note that~\cite{TS} uses
lowercase Latin indices, $T_a$, for spherical tensors,
while~\cite{MP-pert} uses lower case Latin indices for $(t,r)$-plane
tensors and upper case Latin indices, $T_A$, for spherical tensors.
\emph{Formulas given while referring to a given paper will use the notation of
that paper.}

\section{Singularity of `convenient' gauge}\label{sec:gauge}
For canonical quantization of the metric perturbations, it is necessary
to isolate the single physical degree of freedom in the even parity
sector, known in the literature as the \emph{Zerilli} or
\emph{Zerilli-Moncrief} function, the gauge freedom arising from
linearized coordinate transformations must be completely fixed.
Takahashi \& Soda impose what they call a \emph{`convenient' gauge},
where $v=0$, $B=0$ and $K=0$. Further, they solve the constraints
following from the equations of motion and express all remaining
perturbation components in terms of the scalar Zerilli function
$\psi^Z$. I will show that gauge transformations required to enforce the
`convenient' gauge are singular at the horizon. As a consequence, some
remaining non-zero components will be singular at the horizon as well.
This singularity becomes obvious when these components are explicitly
expressed in terms of the Zerilli function in a coordinate system
regular at the horizon.

Let a field be called \emph{regular} at a point $x$ if
it is continuous and smooth in some neighborhood of that point;
otherwise it is called \emph{singular} at $x$. Regularity at $x$ implies
that the components of the tensor field must be continuous, smooth
functions in a neighborhood of $x$, when expressed in any coordinate
system that covers $x$. The converse implication is also true. It is a
sensible mathematical and physical requirement that metric perturbations
be restricted to everywhere regular tensor fields. Note, however, that
regularity in static Schwarzschild coordinates is insufficient to
establish global regularity. Recall that Schwarzschild coordinates
consist of two coordinate charts (exterior, $r>2M$, and interior,
$0<r<2M$), neither of which covers any point on the horizon.  To check
regularity at the horizon, it is necessary and sufficient to check
continuity and smoothness in any coordinate system that does cover the
horizon. We shall use the advanced Eddington-Finkelstein (EF)
coordinates~\cites{MP-pert,PP-lc,MP-coord}; they are regular on the
future horizon, which suffices for our purposes. These $(v,r)$
coordinates are related to the static Schwarzschild $(t,r)$ coordinates as
follows:
\begin{gather}
	r^* = \int\frac{\d{r}}{f} = r+2M\ln\left|\frac{r}{2M}-1\right| , \\
	\begin{aligned}
		v &= t+r^* \\
		r &= r 
	\end{aligned} ,
	\qquad
	\begin{aligned}
		\d{v} &= \d{t}+f^{-1}\d{r} \\
		\d{r} &= \d{r}
	\end{aligned} ,
	\qquad
	\begin{aligned}
		\del_t &= \del_v \\
		\del_r &= f^{-1}\del_v + \del_r
	\end{aligned} .
\end{gather}

\subsection{`Convenient' gauge in EF coordinates}\label{sec:ef}
The gauge chosen by the authors of~\cite{TS} for even-parity
perturbations is termed the \emph{`convenient' gauge}: $K=B=v=0$. In the
notation of~\cite{MP-pert}, the `convenient' gauge is equivalent to
$K=0$, $G=0$, and $j_a (\del_t)^a=0$. A gauge is called \emph{good} if
an arbitrary perturbation $p_{\mu\nu}$ can be transformed by a unique vector
field $\Xi_\mu$ to one that satisfies the gauge condition, $p'_{\mu\nu}
= p_{\mu\nu} - \nabla_\mu \Xi_\nu - \nabla_\nu \Xi_\mu$. I call a gauge
\emph{regular} if $p'_{\mu\nu}$ is everywhere regular whenever
$p_{\mu\nu}$ is everywhere regular. This is equivalent to saying that
there must exist an everywhere regular vector field
$\Xi_\mu=(\xi_a,D_A\xi)$ that implements the required transformation.
Otherwise I call the gauge \emph{singular}.  From appendix~E
of~\cite{MP-pert}, the explicit form of the gauge transformation is
\begin{align} 
\label{gfirst}
h'_{vv} &= h_{vv} -2 \del_v \xi_v 
+ \frac{2M}{r^2} \xi_v + \frac{2Mf}{r^2} \xi_r, \\ 
h'_{vr} &= h_{vr} -\del_r \xi_v 
- \del_v \xi_r - \frac{2M}{r^2} \xi_r, \\
h'_{rr} &= h_{rr} -2 \del_r \xi_r, \\
j'_v &= j_v -\del_v \xi - \xi_v, \\ 
j'_r &= j_r -\del_r \xi - \xi_r 
+ \frac{2}{r} \xi, \\
K' &= K -\frac{2f}{r} \xi_r - \frac{2}{r} \xi_v 
+ \frac{\ell(\ell+1)}{r^2} \xi, \\ 
\label{glast}
G' &= G -\frac{2}{r^2} \xi. 
\end{align}     
The conditions $G'=0$ and $j'_a (\del_t)^a = j'_a (\del_v)^a = j'_v = 0$
are easily obtained by setting $\xi = r^2 G/2$ and $\xi_v = j_v - r^2
\del_v G/2$, independent of any other requirements. On
the other hand, setting $K'=0$ requires
\begin{gather}
	\xi_r = \frac{r}{2f}\left(K - \frac{2}{r}j_v + r\del_vG +
		\frac{\ell(\ell+1)}{2} G\right).
\end{gather}
Note that $f(r=2M)=0$, therefore, since $K$, $G$ and $j_v$ were assumed
to be arbitrary regular functions, $\xi_r$ must be singular at $r=2M$,
the future horizon, which is covered by our choice of advanced EF
coordinates. Hence, the `convenient' gauge is necessarily singular.
Moreover, this singularity appears explicitly in the $j'_r$, $h'_{rr}$
and $h'_{vr}$ components of the metric perturbation, though not in the
$h'_{vv}$ one.

Note that the above result would have been difficult, though not
impossible, to obtain directly in static Schwarzschild coordinates.
Since no point of the horizon is covered by $(t,r)$ coordinates, even
regular tensor fields may have components that diverge as powers of
$1/f$ as $r\to2M$, though such divergences will have a specific
structure. For instance, this structure can be identified by
transforming a regular tensor field from EF to Schwarzschild coordinates.
Thus, the regularity of a tensor fields could be checked in static
coordinates by examining the structure of the divergences of its
components as $r\to2M$, though with some effort. In fact, a variant of
this technique was unsuccessfully, used in section~4.4 of~\cite{TS}.
Unfortunately, the authors ultimately failed to notice the singular
nature of their gauge choice.  Also note that the same procedure can be
used to check that the standard \emph{Regge-Wheeler gauge}%
	\footnote{Referred to as \emph{Zerilli gauge} in~\cite{TS}.} %
($j_a=0$, $G=0$) or the more recently proposed \emph{light-cone gauge}
of Preston~\& Poisson~\cite{PP-lc} are both regular.

\subsection{Zerilli function}
After imposing the `convenient' gauge, the authors of \cite{TS} proceed
to solve the constraints among the remaining independent even-parity
components of $h_{\mu\nu}$, isolate the single physical degree of
freedom (the Zerilli function $\psi^Z$), and express $h_{\mu\nu}$ in
terms of $\psi^Z$. The Zerilli function is quantized as a normal scalar
field. The expression for $h_{\mu\nu}$ in terms of $\psi^Z$ is necessary
to obtain the correct (lowest non-trivial order) coupling between
$\psi^Z$ and a Klein-Gordon field $\phi$. This coupling is obtained from
the cubic term in the expansion of the standard, massless
Einstein-Hilbert-Klein-Gordon Lagrangian.

The derivation of the explicit expressions for $h_{\mu\nu}$ in terms of
$\psi^Z$ in~\cite{TS} are fairly involved and specific to static
Schwarzschild coordinates. However, given that the
Zerilli function is already known in gauge independent form,
cf equation~(4.23) of~\cite{MP-pert}, these expressions may be obtained
algorithmically in any coordinate system, if the gauge-fixed equations
of motion in that coordinate system are given. This opens the
possibility of using computer algebra software to reduce the manual
labor necessary to reproduce their derivation.

For instance, the equations of motion for the even-parity metric
perturbations in EF coordinates are given in appendix~E
of~\cite{MP-pert}. In `convenient' gauge they reduce to the following.
First, we write down the so-called gauge invariant combinations:
\begin{align} 
\tilde{h}_{vv} &= h_{vv} + \frac{2Mf}{r^2} j_r , \\
\tilde{h}_{vr} &= h_{vr}  
- \del_v j_r
- \frac{2M}{r^2} j_r , \\ 
\tilde{h}_{rr} &= h_{rr} 
- 2 \del_r j_r , \\
\tilde{K} &= - \frac{2f}{r} j_r .
\end{align} 
In terms of them, the equations of motion for the metric perturbations,
where for brevity we use $\lambda=\ell(\ell+1)$ and $\mu=(\ell-1)(\ell+2)$,
are
\begin{align} 
\label{EE1}
0 &= -\del_r^2 \tilde{K} 
- \frac{2}{r} \del_r \tilde{K} 
- \frac{1}{r} \del_v \tilde{h}_{rr} 
+ \frac{f}{r} \del_r \tilde{h}_{rr}
+ \frac{2}{r} \del_r \tilde{h}_{vr} 
+ \frac{\lambda r + 4M}{2r^3} \tilde{h}_{rr}, \\ 
0 &= \del_v\del_r \tilde{K} 
+ \frac{2}{r} \del_v \tilde{K} 
+ \frac{r-M}{r^2} \del_r \tilde{K} 
- \frac{f}{r} \del_v \tilde{h}_{rr} 
- \frac{1}{r} \del_r \tilde{h}_{vv}
\\ \notag & \mbox{} \quad
- \frac{1}{r^2} \tilde{h}_{vv} - \frac{\lambda+4}{2r^2} \tilde{h}_{vr} 
- \frac{f}{r^2} \tilde{h}_{rr} - \frac{\mu}{2r^2} \tilde{K}, \\ 
0 &= -\del_v^2 \tilde{K} 
+ \frac{r-M}{r^2} \del_v \tilde{K} 
+ \frac{(r-M)f}{r^2} \del_r \tilde{K}
+ \frac{1}{r} \del_v \tilde{h}_{vv}  
+ \frac{2f}{r} \del_v \tilde{h}_{vr} 
- \frac{f}{r} \del_r \tilde{h}_{vv}  
\\ \notag & \mbox{} \quad
+ \frac{\mu r + 4M}{2r^3} \tilde{h}_{vv}
- \frac{2f}{r^2} \tilde{h}_{vr}  
- \frac{f^2}{r^2} \tilde{h}_{rr} 
- \frac{\mu f}{2r^2} \tilde{K}, \\ 
0 &= \del_v \tilde{h}_{rr}
- \del_r \tilde{h}_{vr} 
- \del_r \tilde{K} 
+ \frac{2}{r} \tilde{h}_{vr} 
+ \frac{r-M}{r^2} \tilde{h}_{rr}, \\
0 &= -\del_v \tilde{h}_{vr}
+ \del_r \tilde{h}_{vv} 
- \del_v \tilde{K} 
- f \del_r \tilde{K} 
+ \frac{2(r-M)}{r^2} \tilde{h}_{vr}
+ \frac{(r-M)f}{r^2} \tilde{h}_{rr}, \\
0 &= -\del_v^2 \tilde{h}_{rr} 
+ 2 \del_v\del_r \tilde{h}_{vr} 
- \del_r^2 \tilde{h}_{vv} 
+ 2 \del_v\del_r \tilde{K} 
+ f \del_r^2 \tilde{K} 
- \frac{r-M}{r^2} \del_v \tilde{h}_{rr} 
+ \frac{2}{r} \del_v \tilde{K}
\\ \notag & \mbox{} \quad
- \frac{2}{r} \del_r \tilde{h}_{vv} 
- \frac{2(r-M)}{r^2} \del_r \tilde{h}_{vr}  
- \frac{(r-M)f}{r^2} \del_r \tilde{h}_{rr} 
+ \frac{2(r-M)}{r^2} \del_r \tilde{K}  
- \frac{\lambda}{r^2} \tilde{h}_{vr} 
\\ \notag & \mbox{} \quad
- \frac{\lambda r^2-2\mu Mr-4M^2}{2r^4} \tilde{h}_{rr}, \\ 
0 &= -2 \tilde{h}_{vr} - f \tilde{h}_{rr}. 
\end{align} 
The Zerilli function, from appendix~E of~\cite{MP-pert}, is given by
\begin{align}
\label{Zerilli}
	\Psi &= \frac{2r}{\lambda}\left[\tilde{K}
		+\frac{2}{\Lambda}(\tilde{h}_{rr}-r\del_r\tilde{K})\right], \\
	\Lambda &= (\ell-1)(\ell+2)+\frac{6M}{r}=\mu+\frac{6M}{r},
	\quad \ell\ge 2.
\end{align}
The restriction on the spherical harmonic index stems from the fact that
the $\ell=0,1$ modes are not dynamical. The denominator $\Lambda$ is
clearly non-vanishing either in the exterior or interior black hole
regions. Note that the above expression may differ by an
$\ell$-dependent multiplicative constant from $\psi^Z$ defined
in~\cite{TS}, which is immaterial for the purposes of this discussion.

The algorithm%
	\footnote{This algorithm is inspired by the study of formal
	integrability of partial differential equations
	equations~\cites{ST1,S2}. It is not too difficult to see how the
	simplified version presented here is equivalent to solving the
	constraint equations `by hand.'} %
for expressing $p_{\mu\nu}$ in terms of $\Psi$ consists
of the following steps. Recall that each of the above equations is
linear in the components of $p_{\mu\nu}$ and their partial derivatives.
Also, let $P_n$ denote the set of all $n$-th partial derivatives of the
$p_{\mu\nu}$ components $h_{vv}$, $h_{vr}$, $h_{rr}$, and $j_r$.
Note that we are not including $\Psi$ or its derivatives in $P_n$.
\begin{enumerate}
	\item \textbf{Initialization}
	Let $E$ be a list of expressions whose vanishing is equivalent to
	equations~\eqref{EE1} to~\eqref{Zerilli}, e.g.,\ the right hand sides of
	those equations. Further, divide this list into subsets $E_n$, each
	containing no more than $n$ partial derivatives acting on the components
	of $p_{\mu\nu}$ (that is, variables from $P_0$ up to $P_n$ only).
	Lastly, define $E_{-1}$ be the subset of expressions that depend on
	$\Psi$ and its derivatives only; it starts out empty.
	\item \textbf{Iteration}
	Repeat for $n=1$, $0$, and $-1$, in that order:
	Apply $\del_v$ and $\del_r$ to each
	element of $E_n$ and collect the results in $E'_{n+1}$. Using linear
	operations, eliminate the variables $P_{n+1}$ (being the highest order
	derivatives) from $E_{n+1}\cup E'_{n+1}$.
	Replace $E_{n+1}$ by the eliminated expressions and add the remaining
	independent expressions to $E_n$, which is possible since the
	remaining expressions will have no more than $n$ derivatives acting on
	each component of $p_{\mu\nu}$.
	\item \textbf{Termination}
	Iterate step~2 until the number of independent expressions in $E_0$ is
	the same as the number of variables in $P_0$. Optionally, keep
	iterating until $E_{-1}$ is non-empty.
	\item \textbf{Explicit Solution}
	Set each expression in $E_0$ to zero and solve the resulting linear
	equations for the variables in $P_0$. Each of the $p_{\mu\nu}$
	components will then be explicitly expressed in terms of $\Psi$ and
	its derivatives.
\end{enumerate}
If $E_{-1}$ is non-empty, then setting each of its elements to zero is
equivalent to the explicit equation of motion for $\Psi$. This algorithm
is not guaranteed to terminate for an arbitrary set of partial
differential equations with constraints (though a generalized version of
it is guaranteed to terminate under fairly general conditions).
However, if it is known to terminate for a set of partial differential
equations expressed in one coordinate system, then it will terminate for
the same set of equations expressed in any other coordinate system. The
results of section~2.1 of~\cite{TS} essentially show that the algorithm
terminates for equations~\eqref{EE1} to~\eqref{Zerilli}, when expressed
in static Schwarzschild coordinates.

Applying this algorithm, we can obtain explicit expressions for the
non-zero components of $p_{\mu\nu}$ in EF coordinates. The
corresponding expressions in static Schwarzschild coordinates can be
obtained by applying the same algorithm to the equations of motion
explicitly given in appendix~C of~\cite{MP-pert} or by applying the
usual coordinate transformation rules of tensor calculus. The results
agree with section~2.1 of~\cite{TS}.

In these explicit expressions,
the singularity of the `convenient' gauge is apparent
from the presence of terms proportional to inverse powers of $1/f$ in
$j_r$, $h_{rr}$ and $h_{vr}$, which diverge as $r\to2M$, presuming that
$\Psi$ is itself regular at the horizon. These divergences can be
removed by the following explicit (singular) gauge transformation:
$\xi=0$, $\xi_v=0$, and 
\begin{equation}\label{expl-gauge}
	\xi_r = - \frac{M}{2\lambda}
		\frac{(4M\del_v\Psi+\lambda\Psi)}{rf} .
\end{equation}
After this gauge transformation, the explicit expressions for the
components of the metric perturbation become
\begin{align}
	h_{vv} &=
- \frac{2 \mu \Psi M ( \lambda + 1 )^2}{3 \Lambda^2 r^2}
- \frac{(\lambda+1)[12 \del_r \Psi M ( \lambda + 1 ) + \mu \Psi ( \lambda - 8 ) + 36 \del_v \Psi M]}{18 \Lambda r} \\
\notag & \quad {}
- \frac{\Psi M^2 \lambda + 4 \del_v \Psi M^3}{r^3}
- \frac{\Psi M (\lambda - 8)}{6 r^2}
+ \frac{\mu \Psi \left( \lambda + 4 \right) + 12 \mu \left( \del_r \Psi \right) M}{18 r} \\
\notag & \quad {}
+ r \del_v^2 \Psi + \del_v \Psi + \del_r \Psi \\
	h_{vr} &=
\frac{2 \mu \Psi M \left( \lambda + 1 \right)}{\Lambda^2 r^2}
- \frac{\mu \Psi \left( \lambda + 4 \right) - 12 \left( \del_r \Psi \right) M \left( \lambda + 1 \right) - 18 \left( \del_v \Psi \right) M}{6 \Lambda r} \\
\notag & \quad {}
- \frac{\Psi M \lambda + 4 \left( \del_v \Psi \right) M^2}{2 r^2}
- \frac{3 \left( \del_v \Psi \right) M + \mu \Psi}{3 r} \\
\notag & \quad {}
- \frac{\lambda \del_v \Psi + 2 \left( \del_v^2 \Psi \right)  \left( 2 M + r \right) - 2 \left( \del_v\del_r \Psi \right) r + 4 \left( \del_r \Psi \right)}{4} \\
	h_{rr} &=
- \frac{6 \mu \Psi M}{\Lambda^2 r^2}
- \frac{6 \left( \del_r \Psi \right) M}{\Lambda r}
- \frac{2 (\del_v\del_r \Psi) \left( 2 M + r \right) + 2 \left( \del_v \Psi \right) + \mu \left( \del_r \Psi \right)}{2}, \\
	j_r &=
\frac{3 \Psi M}{\Lambda r}
- \frac{\Psi \lambda + 2 \left( \del_v \Psi \right)  \left( 2 M + r \right) + 2 \left( \del_r \Psi \right) r}{4} .
\end{align}
These expressions are manifestly regular at the horizon.
Finally, the explicit equation of motion for $\Psi$ is
\begin{equation}
	2 \del_v\del_r \Psi + f \del_r^2 \Psi + f' \del_r \Psi
	=
	\frac{1}{\Lambda^2}\left[
		\mu^2\left(\frac{\mu+2}{r^2}+\frac{6M}{r^3}\right)
		+\frac{36M^2}{r^4}\left(\mu+\frac{2M}{r}\right)\right] \Psi,
\end{equation}
where the left hand side of the above equation is simply the
d'Alambertian, $\square \Psi$, on the $(t,r)$-plane. This is the well
known equation of motion for the Zerilli function, derived both
in~\cite{TS} and~\cite{MP-pert}.

\section{Interaction and divergences}\label{sec:loop}
\begin{figure}
\begin{fmffile}{scalgrav}%
	\newsavebox{\loopbox}%
	\savebox{\loopbox}{%
	\begin{fmfgraph*}(50,30)
		\fmfleft{i} \fmfright{o}
		\fmf{plain,label=$\omega$,tension=4}{i,v1}
		\fmf{plain,label=$\omega$,tension=4}{v2,o}
		\fmf{plain,right}{v1,v2}
		\fmf{wiggly,left=1.1}{v1,v2}
	\end{fmfgraph*}%
	}%
	\newsavebox{\cutloopbox}%
	\savebox{\cutloopbox}{\fmfframe(1,2)(4,2){%
	\begin{fmfgraph*}(25,30)
		\fmfleft{i} \fmfright{so,go}
		\fmfv{l=$\omega'$,l.a=0}{so}
		\fmfv{l=$\omega''$,l.a=0}{go}
		\fmf{plain,label=$\omega$,tension=3}{i,v1}
		\fmf{plain,right=.4}{v1,so}
		\fmf{wiggly,left=.4}{v1,go}
	\end{fmfgraph*}%
	}}%
	\begin{center}
	\begin{Large}
	$\displaystyle
		\raisebox{-.5\ht\loopbox+.5ex}{\usebox{\loopbox}}
	~~ \sim ~ \sum_{\omega',\omega''}
		\left|\raisebox{-.5\ht\cutloopbox+.5ex}{\usebox{\cutloopbox}}\right|^2
	$
	\end{Large}
	\end{center}
	\caption{Lowest order correction to the scalar (straight lines)
	$2$-point function from interaction with dynamical gravitons (wavy
	lines). The $\omega$ labels are short-hand for the mode indices and
	the sums may involve integrals over continuous labels. The loop diagram
	may be evaluated, using an optical theorem-like identity,
	by cutting it into two tree diagrams.}
	\label{loop-diag}
\end{fmffile}
\end{figure}
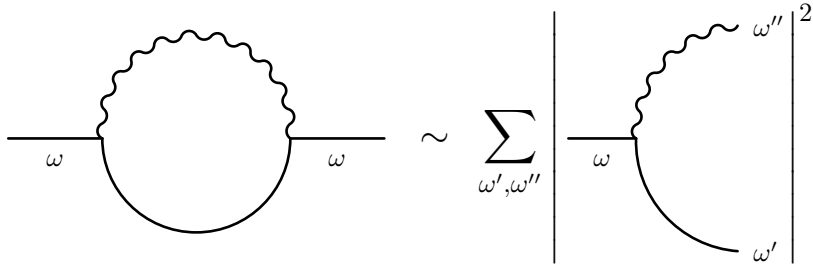

Once both the scalar field and the metric perturbations are quantized,
and their cubic coupling is explicitly derived, the calculation
in~\cite{TS} proceeds as follows (though this logic is only implicit in
its technical details). The lowest order correction to the scalar
2-point function is computed, which amounts to taking into account the
single loop diagram shown in figure~\ref{loop-diag}. This diagram is
evaluated using an optical theorem-like identity, which is also
schematically illustrated in figure. The correction to the scalar
Hawking radiation spectrum is encoded in this correction to the 2-point
function.

On a Minkowski background, this kind of calculation is most conveniently
performed in momentum space, where each leg of the interaction vertex
appearing in figure~\ref{loop-diag} is labelled by an on-shell
4-momentum $k$ and a tensor polarization index $j$ (if any), while the
vertex factor is proportional to the triple mode function overlap
integral
\begin{align}
	V_{k,k',k''} &= \int\d^4x\,\phi^0_k(x)\phi^0_{k'}(x)\phi^0_{k''}(x) \\
		&= \int\d^4x\,\exp[-i(k+k'+k'')_\mu x^\mu]
		\sim \delta(k+k'+k'') .
\end{align}
It is apparent, since each mode function is everywhere regular,
that the integrand defining the vertex factor is \emph{locally
integrable}, that is, its integral over any bounded region exists and is
finite. It is also apparent that, nonetheless, $V_{k,k',k''}$ is a
distribution, which follows from the global convergence properties of
the above integral. Since these properties rely only on the local
regularity of mode functions, they are expected to hold in curved
space-times as well.

On the black hole background, it is most convenient to perform
diagrammatic calculations in angular momentum-frequency space. Each line
of a diagram is then labelled by a frequency $\omega$, a pair of
spherical harmonics indices $\ell, m$, and a tensor polarization index
$j$. Roughly speaking, the vertex factor is once again proportional to
the triple mode function overlap integral
\begin{equation}\label{bh-3mod}
	V_{\omega\ell,\omega'\ell',\omega''\ell''}
	= \int w(r)\,
		\phi^{\pm}_{\omega\ell}(t,r)
		\phi^{\pm}_{\omega'\ell'}(t,r)
		\phi^{\pm}_{\omega''\ell''}(t,r) ,
\end{equation}
where $w(r)$ is a 2-form on the $(t,r)$-plane, which takes into account
the invariant volume measure and $r$-dependent coefficients that come
from the expression for $h_{\mu\nu}$ in terms of $\psi^Z$. Our
discussion up to this point shows that, in a regular gauge, each of the
$w$, $\phi^{\pm}_{\omega\ell}$, as well as their products, should be
locally integrable, including in the vicinity of the horizon.

More precisely, to take into account all kinds of vertices that couple
the scalar field to metric perturbations, we must also consider
derivative couplings. In that case, some of the $\phi_{\omega\ell}$'s
in~\eqref{bh-3mod} will be acted upon by partial derivatives. If each
$\phi_{\omega\ell}$ is regular, then any scalars made up of its
derivatives will also be regular, therefore the local integrability
argument is unmodified. Note, however, due to the presence of the black
hole horizon, that the $\phi^+_{\omega\ell}$ fail to be regular at the
horizon. They are still bounded, but become highly oscillatory in the
vicinity of the future horizon, with the oscillation phase diverging at
the horizon itself (see section~10.2 of \cite{FN-bh}, for instance).
Hence, derivatives of $\phi^+_{\omega\ell}$ may become unbounded, though
highly oscillatory, in a neighborhood of the horizon. Nonetheless,
despite being unbounded, due to the oscillatory behavior of the
integrands, their integrals should be evaluable in a distributional
sense. Hence, even on a black hole background, and even with derivative
couplings, the integrand in the triple mode function
overlap~\eqref{bh-3mod} should be locally integrable, though perhaps
only distributionally.

\subsection{Divergence in the triple mode function overlap}
The expressions $K^{\omega\ell}_{\omega'\ell';\omega''\ell''}$ (for
even-parity modes) and $H^{\omega\ell}_{\omega'\ell';\omega''\ell''}$
(for odd-parity modes), introduced in equation~(128) of~\cite{TS}, are
closely related to these kinds of triple mode function overlap
integrals. Despite the expectations expressed above, equations~(160) to
(162) of~\cite{TS} show that the integrand defining
$K^{\omega\ell}_{\omega'\ell';\omega''\ell''}$ fails to be locally
integrable. This failure of local integrability can be traced to the
singularity of the `convenient' gauge, which introduces terms of the
form $\int\d{\mathrm{vol}}/f^n$ with $n=1,2$ into the integral
in~\eqref{bh-3mod}, making it diverge in the vicinity of the horizon. In
fact, as follows from the results of section~\ref{sec:gauge}, the only
such locally non-integrable terms that contribute to the
$K^{\omega\ell}_{\omega'\ell';\omega''\ell''}$ are precisely the ones
that can be removed by the explicit gauge
transformation~\eqref{expl-gauge} and hence are pure gauge artifacts. On
the other hand, the integrand defining $V_{\omega,\omega',\omega''}$
would be locally integrable, as expected, in any regular gauge.

Unfortunately, the authors of~\cite{TS} have mistakenly identified these
divergent contributions to
$K^{\omega\ell}_{\omega'\ell';\omega''\ell''}$ as the dominant ones,
have arbitrarily regulated them using a principal value prescription in
the radial integral over an interval of size $\eps$ about the horizon,
and have dropped all other terms, including
$H^{\omega\ell}_{\omega'\ell';\omega''\ell''}$. Since this
$\eps$-regulator appears in the multiplicative coefficient of the
spectrum correction in~\eqref{spec-corr}, the only conclusion to be
drawn from the preceding discussion is that the size of the correction
\emph{has not} been correctly estimated and that what \emph{has} been
estimated is a pure gauge artifact. To obtain a reliable estimate, the
triple mode function overlap integrals would have to be analyzed anew,
once they are rewritten in a regular gauge.

\subsection{Divergence in the summations over modes}
The logic outlined at the beginning of this section culminates in
equation~(144) of~\cite{TS}, which expresses the correction to the
Hawking radiation spectrum in terms of the triple mode function
overlaps, $K^{\omega\ell}_{\omega'\ell';\omega''\ell''}$ and
$H^{\omega\ell}_{\omega'\ell';\omega''\ell''}$, the Bogolubov
coefficients relating the $\phi^\pm_{\omega\ell j}$ modes,
$\alpha_{\omega\ell,\omega'\ell'}$ and
$\beta_{\omega\ell,\omega'\ell'}$, and some Clebsch-Gordan coefficients
coming from the integration of products of spherical harmonics, $C^{\ell
m}_{\ell'm';\ell''m''}$. This expression for the correction is
schematically, as illustrated in figure~\ref{loop-diag},
\begin{equation}\label{loop-int}
	\sum_{\omega',\omega''} |V_{\omega,\omega',\omega''}|^2 ,
\end{equation}
where each $\sum_\omega$ compactly represents a combined sum-integral
over all mode indices, also including spherical harmonic and
polarization indices. Once $V_{\omega,\omega',\omega''}$ is estimated,
the outer mode sums are seen to be divergent. In equations~(163)
and~(164) of~\cite{TS}, this divergence is regulated by essentially introducing
lower and upper frequency cutoffs, respectively $1/L$ and
$\omega_{\mathrm{cut}}$, cf~also equation~(82) in~\cite{TS}.

Leaving aside the fact that the estimates of the size of
$V_{\omega,\omega',\omega''}$ cannot be completely trusted due to gauge
artifacts, a divergence in~\eqref{loop-int} is to be expected, as in any
generic 1-loop perturbative calculation. The standard way to deal with
such a divergence is to introduce a local counter-term in the original
Lagrangian density. Since this divergence appears in a correction to the
scalar self-energy, as illustrated in figure~\ref{loop-diag}(a), such a
counter-term would only renormalize the kinetic and mass parts of the
scalar field Lagrangian density. Thus, renormalization would allow the
regulator dependence of the final result to be removed, such that the
coefficient in front of the correction in~\eqref{spec-corr} would not
depend on $\omega_{\mathrm{cut}}$ and $L$.

\section{Discussion}\label{sec:discuss}

The authors of~\cite{TS} have tackled an interesting and challenging
question. Unfortunately, their calculation suffers from a few problems.
The so-called `convenient' gauge chosen by the authors for the
even-parity metric perturbations turns out to be singular, unlike the
standard Regge-Wheeler gauge. The singularities introduced by their
choice of gauge result in spurious divergences, which mask all other
contributions in the vertex factors characterizing the coupling of
metric perturbations to the scalar field. Moreover, another divergence,
corresponding to the expected 1-loop divergence of perturbative quantum
field theory, is not removed via renormalization. Both kinds of
divergences are regulated, introducing arbitrary parameters into the
calculation. As clearly seen in~\eqref{spec-corr}, the final result for
the correction to the spectrum of scalar Hawking radiation depends on
regulators $\eps$, $\omega_{\mathrm{cut}}$ and $L$. Their presence,
makes the given estimate for the size of the correction unreliable.

Despite these problems, the authors have successfully addressed major,
necessary parts of the calculation: (a) quantization of a scalar field
and metric perturbations in a black hole background (via gauge fixing
and explicit reduction to physical degrees of freedom), (b) explicit
evaluation of the cubic scalar-graviton coupling (modulo gauge issues),
(c) estimation of the in-out mode Bogolubov coefficients via a geometric
optics approximation and (d) an explicit expression for the spectrum
correction in terms of Bogolubov coefficients, triple mode function
overlaps and Clebsch-Gordan coefficients. It would suffice only minor
modifications and a careful application of standard quantum field
theoretic techniques to complete this calculation and obtain a definite,
parameter-free estimate for the correction to the spectrum of scalar
Hawking radiation. Moreover, the same techniques are readily applicable
to the problem of perturbative back reaction of Hawking radiation on the
quantum geometry of the black hole.

\section*{Acknowledgments}
The author would like to thank Tomohiro Takahashi, Erik Plauschinn,
Alessandro Torielli, Paul Reska and Tomislav Prokopec for helpful
discussions.  The author was supported by a Postdoctoral Fellowship from
the National Science and Engineering Research Council (NSERC) of Canada.

\end{document}